\documentclass[]{JHEP}
\usepackage{psfig}
\newcommand{\newc}{\newcommand}
\newc{\ra}{\rightarrow}
\newc{\lam}{\lambda}
\newc{\eps}{\epsilon}
\newc{\half}{\frac{1}{2}}
\newc{\third}{\frac{1}{3}}
\newc{\fourth}{\frac{1}{4}}
\newc{\eighth}{\frac{1}{8}}
\newc{\gev}{\mbox{~GeV}}
\newc{\lra}{\leftrightarrow}
\newc{\Dslash}{\not\!\! D}
\newc{\sg}{{\cal G}}
\newc{\etal}{{\it et al.}\ }
\newc{\Hbar}{{\bar H}}
\newc{\hhbar}{{\overline h}}
\newc{\Ubar}{{\bar U}}
\newc{\Dbar}{{\bar D}}
\newc{\Ebar}{{\bar E}}
\newc{\eg}{{\it e.g.}\ }
\newc{\ie}{{\it i.e.}\ }
\newc{\nonum}{\nonumber}
\newc{\kap}{\kappa}
\newc{\Dt}{\frac{d}{dt}}
\newc{\rpv}{$\not\!\!R_p$}
\newc{\lfv}{$\not\!\!L$}
\newc{\mpl}{$M_{Pl}$\ }
\newc{\mx}{$M_X$\ }
\newc{\tev}{\mbox{~TeV}}
\newc{\sect}[1]{(\ref{sec:#1}}
\newc{\nonr}{\nonumber}
\newc{\lab}[1]{\label{eq:#1}}
\newc{\Lam}{{\Lambda}}
\newc{\ltau}{\lambda_\tau}
\newc{\lt}{\lambda_t}
\newc{\lb}{\lambda_b}
\newc{\lae}{{\Lam}_E}
\newc{\lad}{{\Lam}_D}
\newc{\lau}{{\Lam}_U}
\newc{\lame}[1]{{\Lam}_{E^{#1}}}
\newc{\lamhe}[1]{{\h}_{E^{#1}}}
\newc{\lamhed}[1]{{\h}_{E^{#1}}^\dagger}
\newc{\lamhd}[1]{{\h}_{D^{#1}}}
\newc{\lamhdd}[1]{{\h}_{D^{#1}}^\dagger}
\newc{\lamhu}[1]{{\h}_{U^{#1}}}
\newc{\lamhud}[1]{{\h}_{U^{#1}}^\dagger}
\newc{\lamd}[1]{{\Lam}_{D^{#1}}}
\newc{\lamu}[1]{{\Lam}_{U^{#1}}}
\newc{\lamet}[1]{{\Lam}_{E^{#1}}^T}
\newc{\lamdt}[1]{{\Lam}_{D^{#1}}^T}
\newc{\lamut}[1]{{\Lam}_{U^{#1}}^T}
\newc{\lames}[1]{{\Lam}_{E^{#1}}^*}
\newc{\lamds}[1]{{\Lam}_{D^{#1}}^*}
\newc{\lamus}[1]{{\Lam}_{U^{#1}}^*}
\newc{\lamed}[1]{{\Lam}_{E^{#1}}^\dagg}
\newc{\lamdd}[1]{{\Lam}_{D^{#1}}^\dagg}
\newc{\lamud}[1]{{\Lam}_{U^{#1}}^\dagg}
\newc{\Y}{{Y}}
\newc{\h}{{h}}
\newc{\ye}{{\Y}_E}
\newc{\he}{{\h}_E}
\newc{\hed}{{\h}_E^\dagger}
\newc{\yd}{{\Y}_D}
\newc{\hd}{{\h}_D}
\newc{\hdd}{{\h}_D^\dagger}
\newc{\yu}{{\Y}_U}
\newc{\hu}{{\h}_U}
\newc{\hud}{{\h}_U^\dagger}
\newc{\yes}{{\Y}_E^*}
\newc{\yds}{{\Y}_D^*}
\newc{\yus}{{\Y}_U^*}
\newc{\yet}{{\Y}_E^T}
\newc{\ydt}{{\Y}_D^T}
\newc{\yut}{{\Y}_U^T}
\newc{\yed}{{\Y}_E^\dagg}
\newc{\ydd}{{\Y}_D^\dagg}
\newc{\yud}{{\Y}_U^\dagg}
\newc{\dagg}{\dagger}
\newc{\lp}{\left(}
\newc{\rp}{\right)}
\newc{\inv}{\frac{1}{16\pi^2}}
\newc{\invsq}{\frac{1}{(16\pi^2)^2}}
\newc{\ggam}[2]{\Gamma_{#2}^{#1}}
\newc{\yukgam}[2]{\inv \gamma_{#1}^{(1){#2}}+\invsq\gamma_{{#1}}^{(2){#2}}}
\newc{\susyunif}{ohman,nirpaul,marcelacarlos,susyunif}
\newc{\lsim}{\stackrel{<}{\sim}}
\newc{\gsim}{\stackrel{>}{\sim}}
\newc{\Tr}{{~\rm Tr}}
\newc{\me}{{(\bf m_{\tilde{E}}}^2)}
\newc{\mh}[1]{m_{H_{#1}}^2}
\newc{\ml}{{(\bf m_{\tilde{L}}}^2)}
\newc{\md}{{(\bf m_{\tilde{D}}}^2)}
\newc{\mup}{{(\bf m_{\tilde{U}}}^2)}
\newc{\mq}{{(\bf m_{\tilde{Q}}}^2)}
\newc{\mlh}[1]{({\bf m}_{\tilde{L}_{#1} H_1}^2)}
\newc{\mhl}[1]{({\bf m}_{H_1 \tilde{L}_{#1}}^2)}
\newc{\maux}{M_{\rm aux}}
\newc{\ntw}{\longmapsto\hspace{-0.22in}/ \hspace{0.1in}}

\preprint{DAMTP-2000-14}

\title{R-parity violating anomaly
mediated supersymmetry breaking}

\author{B.C.  Allanach \\ DAMTP, Wilberforce Rd, Cambridge CB3 0WA, UK}

\author{A. Dedes \\ Rutherford Appleton Laboratory, Chilton, Didcot, Oxon,
OX11 0QX, UK} 


\abstract{
We propose a new scenario that solves the slepton negative mass squared
problem of the minimal supersymmetric standard model with anomaly
mediated supersymmetry breaking. The solution is achieved by including
three trilinear R-parity violating operators in the superpotential. The soft
supersymmetry breaking terms satisfy renormalisation group invariant relations
in terms of supersymmetric couplings and the overall supersymmetry breaking
mass scale. Flavour changing neutral currents can be naturally highly
suppressed. A specific model predicts $\tan \beta=4.2 \pm 1.0$. Excluding
sleptons, the supersymmetric particle spectrum then depends upon two remaining
free parameters. In the case of the R-parity violating couplings set at their
quasi-fixed points at a supersymmetric GUT scale, the whole sparticle spectrum
approximately depends upon only one free parameter. Imposing experimental
limits leads to a constrained and distinctive phenomenology. The lightest
CP-even Higgs of mass $m_h=118$ GeV would be seen at the Tevatron. All
sparticles and heavy Higgs would evade detection except for the lightest
charginos and neutralinos, whose distinctive leptonic decays would be seen at
the LHC. 
}

\begin{document}


\newpage
\section{Introduction}

Low energy supersymmetry remains the most promising known perturbative solution
to the gauge hierarchy problem that afflicts the Standard Model. It is clear
from current data however, that for supersymmetry to be present in nature, it
must be broken. Three phenomenologically distinct mechanisms for translating
supersymmetry breaking 
from a hidden sector to the observable sector are currently recognised:
tree-level gravity, gauge or anomaly mediation.
The last mediator has received relatively little attention, and it is upon
this mechanism that we focus the attention of this letter.

Anomaly mediated supersymmetry breaking (AMSB) in the minimal supersymmetric
standard model (MSSM) provides a potential solution to the supersymmetric
(SUSY) 
flavour problem~\cite{rs}. This is a problem of many supergravity theories in
which squarks and sleptons typically acquire unacceptably large
flavour-changing neutral currents (FCNCs) through flavour mixings in their
mass matrices.
In AMSB, 
SUSY breaking is assumed to take place in
a hidden (``sequestered'') 
sector.
A re-scaling anomaly in the super-Weyl conformal transformation transmits the
SUSY breaking to the observable sector.
It was suggested that the MSSM superfields be confined to a 3-brane in a
higher dimensional bulk space-time, with the sequestered sector
separated by the extra dimension from the visible sector brane.
If direct Kahler couplings between the sequestered and visible sectors are
suppressed (as is the case in the above geometrical set-up), these SUSY
breaking terms can be the dominant forms of SUSY 
breaking in the visible sector.
This scenario produces a supersymmetric spectrum
dependent upon only three
unknown parameters, an overall supersymmetric breaking mass scale and the MSSM
Higgs potential parameters $\mu$ and $B$.
For example, the AMSB mass squared values for scalar components of  chiral
matter supermultiplets are given by~\cite{rs},
\begin{equation}
(m^2)_{\Phi_i}^{\Phi_j}|_{AM}  = -\frac{1}{4}~\maux^2
\biggl [ \mu \frac{d^2 }{d \mu^2}{\ln Z}_i^j \biggr ],
\end{equation}
where $\mu$ denotes
the t'Hooft renormalization scale and $Z^i_j$ is the matter field wave function
of the superfield $\Phi_i$. $\maux$ is the vacuum expectation value of a
compensator superfield~\cite{rs}, and sets the overall mass scale for visible
sector SUSY breaking.
Defining  
\begin{equation}
\Gamma_{\Phi_i}^{\Phi_j} \equiv -1/2 \frac{d(\ln
Z^i_j)}{d \ln \mu}, 
\end{equation}
we may write
\begin{equation}
(m^2)_{\Phi_i}^{\Phi_j}|_{AM} = 
\frac{1}{2}~\maux^2
\biggl [\beta(Y)\frac{\partial}{\partial Y} {\Gamma}_{\Phi_i}^{\Phi_j}  +
\beta(g)\frac{\partial }{\partial g} {\Gamma}_{\Phi_i}^{\Phi_j}\biggr ]
\label{general}
\end{equation}
summed over all Yukawa couplings $Y$ and gauge couplings $g$. 
$\beta(x)$ represents
the beta function $d x / d \ln \mu$ of coupling $x$. 
An interesting fact is that the AMSB soft terms are valid to
all orders in perturbation theory~\cite{allorders}.
For scalars of the first two families, Yukawa couplings can be safely
neglected and so the dominant terms in eq.~(\ref{general}) are those
proportional to the gauge couplings. These, being family universal, 
highly suppress the most problematic FCNC processes and thus solve the
SUSY flavour problem.

The trilinear soft term $A_Y$ corresponding to Yukawa coupling $Y$ is given by~\cite{feng}
\begin{equation}
Y A_{Y} = - \beta(Y) \maux,
\end{equation}
and the gaugino mass $M_g$ associated with each gauge group of coupling $g$
is~\cite{feng}
\begin{equation}
g M_g = \beta(g) \maux. \label{general2}
\end{equation}

One particularly desirable feature is that
eqs.~(\ref{general}-\ref{general2}) are
renormalisation invariant~\cite{rs}, provided that there are no massive
parameters present in the superpotential (for example from bilinear terms).
This means that AMSB  has the advantage over gravity and gauge
mediation that the low energy
phenomenology does not depend upon the renormalisation of parameters at high
scales, where unknown corrections from new physics apply.
Unfortunately in the minimal version of the AMSB MSSM, the sleptons have
negative mass squared values, indicating that the true vacuum
state of the model is not the desired electroweak one.
There have been several
successful attempts to fix this 
problem, for example
positive bulk Standard Model singlet contributions to
the scalar masses~\cite{rs}, non-decoupling effects~\cite{kss}, heavy
vector-like messengers coupled to light modulus fields~\cite{pr}
have all been proposed. 
All of the above models have spoiled the desirable renormalisation
group invariant feature, rendering the theories potentially sensitive to
unknown ultra-violet effects.

In a recent paper~\cite{FItermsmodel}, a model which solves the slepton
problem in 
AMSB with SUSY breaking Fayet-Iliopoulos D-terms of an additional U(1) gauge
symmetry was presented. Extra Standard-Model gauge singlet chiral superfields
are added to the MSSM\@.
The model has the advantage of soft terms that do not
depend upon the ultra-violet physics~\cite{FItermsmodel},\cite{Dterms}.
The model of ref.~\cite{zmie} extends the MSSM by 3 extra Higgs doublets, a
vector-like pair of extra triplets and 4 new singlets near the weak
scale. Large Yukawa couplings
between the extra Higgs and MSSM leptons provide additional positive
contributions to the slepton mass squareds in eq.~(\ref{general}), while
preserving renormalisation group invariance of the mass relations.
We note that the above attempts to solve the AMSB slepton problem have all
had $R$-parity invariance as a feature. 

Here, we make a new proposal which preserves the renormalisation invariance of
the AMSB supersymmetry breaking mass relations and requires no superfields
additional to those in the MSSM coupling directly to the visible sector.
By considering a subset of
trilinear R-parity violating (\rpv) operators in the superpotential, we
change the wave-function renormalisation of the sparticles, in particular
providing new
positive contributions to the slepton mass squared values. 
Throughout this work we will
assume the dogma of minimality with respect to solving the  slepton
problem in AMSB, for brevity and simplicity.

First, in section~\ref{sec:rescue}, we will classify models that
simultaneously solve
the MSSM AMSB tachyonic slepton problem while not generating dangerous lepton
flavour violating operators.
There emerges a scenario with 3 non-negligible lepton number violating
(\lfv)-operators only.
In section~\ref{sec:soft}, we then present the slepton masses in terms of the
supersymmetric couplings explicitly and briefly discuss the other soft
breaking terms. All other soft masses are equivalent to the $R_p$ conserving
AMSB MSSM to one-loop order. In section~\ref{sec:cons}, we impose constraints
upon the model, the most restrictive being from lepton non-universality.
For the \lfv-contributions to be sufficient to raise all slepton mass squared
values above
zero, we require some \lfv-couplings of order 1. We demonstrate that this is
not in conflict with current data if the scalar sparticles are
rather heavy, above 1.2 TeV. 
We also present the sparticle spectra. In section~\ref{sec:coll}, some
implications for collider searches at the Tevatron and LHC are presented.
Finally, we summarise the main features of the model, reviewing its successful
features and noting possible future work in section~\ref{sec:conc}.

\section{Rescuing the AMSB MSSM with R-parity violation}
\label{sec:rescue}

We begin with the AMSB MSSM including general trilinear \rpv-operators. 
We then identify the subset of operators which are useful in solving the AMSB
slepton problem.
In
the notation of ref.~\cite{second}, the general trilinear
\rpv-MSSM superpotential is written
%
\begin{eqnarray}
W_3&=&
(\ye)_{ij}  L_i H_1 {\bar E}_j +
(\yd)_{ij} Q_i H_1 {\bar D}_j +
(\yu)_{ij} Q_i H_2 {\bar U}_j
+ \nonumber \\ &&
 \frac{1}{2} \lam_{ijk} L_i L_j{\bar E}_k +
\lam_{ijk}^\prime L_i Q_j {\bar D}_{k} +
\frac{1}{2} \lam_{ijk}^{\prime\prime} {\bar U}_i{\bar
D}_j{\bar D}_k,
\label{superpot}
\end{eqnarray}
where we have suppressed gauge indices, and
$i,j,k,\ldots=1,2,3$ are family indices. 
The anomalous dimensions ${\Gamma}_{\Phi_i}^{\Phi_j}$ relevant for
substitution into eq.~(\ref{general}) for the superpotential $W_3$
have been presented in ref.~\cite{uslot} to two-loop order. Their one-loop
truncation is annexed here to Appendix A for ease of reference.

For example, substituting $\Gamma_{{E_R}_i}^{{E_R}_j}$ 
into 
eq.~(\ref{general}),
we obtain the most problematic
soft mass squared, that of the 
right handed sleptons:
\begin{equation}
16 \pi^2 (m^2_{E_R})^j_i = \frac{1}{2} \maux^2 \left[  2 (\yed)_{jk}
						       \beta(\ye)_{ki}  +
						       \lam_{kmi}
						       \beta(\lam_{kmj})
						       + (i \leftrightarrow j)
- G_E						       \right],
\label{RHsleptons} 
\end{equation}
where $G_E\equiv 396 g_1^4 / (25 \times 16 \pi^2)$. The 
first term on the right hand side
is negligible for selectrons and smuons because it is proportional to the
electron and muon mass respectively. In the R-parity conserving
scenario where all 
$\lam_{ijk}=0$, the last (negative) term therefore forces
the right handed selectrons and smuons to have negative mass squared values.
If $\tan \beta > 40$, the positive contribution from $(\ye)_{33}$ becomes
non-negligible and the right handed stau mass squared may be raised above zero.
However, we immediately see that a positive contribution may be obtained to
$(m_{E_R})^k_k$
from $\lam_{ijk} \neq 0$, and it is this possibility that we
exploit\footnote{Following the minimality dogma, we assume real
\rpv-couplings.}.
So far, the condition of minimality identifies\footnote{We set all \rpv
operators not explicitly mentioned to zero.} the combination
\begin{equation}
 \lam_{jk1}, \lam_{lm2}, \lam_{nq3} \neq 0, \label{model2}
\end{equation}
which provides positive contributions to all three right handed slepton
masses. 
We also observe that $\lam_{ijk}'$, $\lam_{ijk}''$ cannot solve the
problem of negative right handed slepton masses and so we drop them from the
discussion. We will assume in the present models that they are zero, and
indeed this assumption will make it simpler to satisfy stringent empirical
limits upon successful scenarios.

The left-handed sleptons also have negative mass squared values in the usual
$R_p$-conserving AMSB MSSM scenario. Including LLE operators by substituting
$\Gamma_{{L}_i}^{{L}_j}$ into eq.~(\ref{general}) and setting
$\lam^\prime_{ijk}, \lam^{\prime\prime}_{ijk}=0$  for all $i,j,k$,
\begin{equation}
16 \pi^2 (m^2_{L})^j_i = \frac{1}{2} \maux^2 \left[    (\yed)_{jk}
						       \beta(\ye)_{ki}  +
						       \lam_{ikq}
						       \beta(\lam_{jkq})
						       + (i \leftrightarrow j)
						       -G_L
						       \right], 
\label{LHsleptons} 
\end{equation}
where $G_L \equiv (99 g_1^2 + 75 g_2^2)/(25 \times 16 \pi^2)$.
Thus the R-parity conserving scenario (all $\lam_{ijk}=0$)
suffers from negative mass squared values for the left-handed selectron and
smuon (and the stau if $\tan \beta$ is not large), analogous to the right
handed sleptons. A positive contribution to $(m^2_{L})^i_i$ 
results if $\lam_{ijk}\neq 0$, i.e.\ we require
\begin{equation}
 \lam_{1jk}, \lam_{2lm}, \lam_{3nq} \neq 0, \label{model}
\end{equation}
to provide additional positive contributions to all left-handed slepton mass
squareds. 

In order to keep the number of couplings to a minimum, we require that the
same non-zero couplings that render the right-handed slepton masses squared
values positive
in eq.~(\ref{model2}) also provide us with positive left-handed
slepton mass squared values in eq.~(\ref{model}).

We now identify a further constraint upon $\lam_{ijk}$
\begin{equation}
\lambda_{imm}=0, \;\;\; \forall i  \label{s1}
\end{equation}
(no sum on $m$) to avoid the generation of large off-diagonal slepton mass
terms. Such terms would generate an empirically unacceptable  amount of lepton
flavour violation~\cite{masiero}, such as $\mu \rightarrow e \gamma$.
Eq.~(\ref{s1}) also forbids the generation of lepton-Higgs mixing, as can be
seen from eq.~(\ref{LHmix}).
Simultaneously applying the above constraints in eqs.~(\ref{model2}-\ref{s1}) 
leads to 
the unique combination
\begin{equation}
\lambda_{123}, \lambda_{132},
 \lambda_{231} \neq 0.
\label{lle}
\end{equation}
Thus we have identified the \lfv-couplings that will solve the AMSB MSSM
slepton problem without generating lepton flavour violating effects that are
too large. We note that for high $\tan \beta > 40$, we could set
$\lam_{123}=0$ and still have positive mass squared values for the
sleptons. For the moment, we include all three couplings for generality, and
indeed below, we focus on a particular model for which 
$\tan \beta=4.2$, requiring us to include $\lam_{123} \neq 0$.

\section{Soft supersymmetry breaking terms}
\label{sec:soft}

We now discuss the (one loop) equations for the soft supersymmetry breaking
terms. We work in a basis where 
Yukawa couplings apart from $Y_\tau\equiv
(\ye)_{33}$, $Y_t \approx (\yu)_ {33}$,
$Y_b \approx (\yd)_{33}$ (the tau, top and bottom Yukawa couplings
respectively) and 
\rpv-couplings {\em not}\/ discussed above are sub-leading and are therefore
neglected.

\subsection{Slepton masses}

The slepton soft masses are given by
eqs.~(\ref{LHsleptons}),(\ref{RHsleptons})
as 
\begin{eqnarray}
(m^2)_{L_1}^{L_1} &=& \frac{M_{aux}^2}{(16\pi^2)} \biggl [ \lam_{123}
\beta(\lam_{123})+\lam_{132}\beta(\lam_{132}) - \biggl (\frac{3}{10} g_1
\beta(g_1)+\frac{3}{2} g_2\beta(g_2) \biggr ) \biggr ] \\
(m^2)_{L_2}^{L_2} &=& \frac{M_{aux}^2}{(16\pi^2)} \biggl [ \lam_{231}
\beta(\lam_{231})+\lam_{123}\beta(\lam_{123}) - \biggl (\frac{3}{10} g_1
\beta(g_1)+\frac{3}{2} g_2 \beta(g_2) \biggr ) \biggr ] \\
(m^2)_{L_3}^{L_3} &=& \frac{M_{aux}^2}{(16\pi^2)} \biggl [ Y_\tau \beta(Y_\tau)
+ \lam_{132}
\beta(\lam_{132})+\lam_{231}\beta(\lam_{231}) 
- \biggl (\frac{3}{10} g_1
\beta(g_1)+\frac{3}{2} g_2 \beta(g_2) \biggr ) \biggr ] \nonumber \\
(m^2)_{E_1}^{E_1} &=& \frac{M_{aux}^2}{(16\pi^2)} \biggl [ 2\lam_{231}
\beta(\lam_{231}) - \frac{6}{5} g_1 \beta(g_1) \biggr ] \\
(m^2)_{E_2}^{E_2} &=& \frac{M_{aux}^2}{(16\pi^2)} \biggl [ 2\lam_{132}
\beta(\lam_{132}) - \frac{6}{5} g_1 \beta(g_1) \biggr ] \\
(m^2)_{E_3}^{E_3} &=& \frac{M_{aux}^2}{(16\pi^2)} \biggl [ 2 Y_\tau
\beta(Y_\tau) + 2\lam_{123}
\beta(\lam_{123}) - \frac{6}{5} g_1 \beta(g_1) \biggr ] 
\end{eqnarray}
where the $\beta$-functions are given by 
\begin{eqnarray}
\beta(Y_\tau) &=& \frac{Y_\tau}{16\pi^2} \biggl [ 4 Y_\tau^2 +
3 Y_b^2 +2 \lam_{123}^2 + \lam_{132}^2+\lam_{231}^2 - \biggl (
\frac{9}{5} g_1^2 + 3 g_2^2 \biggr ) \biggr ] \label{ytaui}\\
\beta(\lam_{123}) &=& \frac{\lam_{123}}{16\pi^2} \biggl [
2 Y_\tau^2+4 \lam_{123}^2+\lam_{231}^2+\lam_{132}^2 -
\biggl (\frac{9}{5} g_1^2+3 g_2^2 \biggr ) \biggr ] \\
\beta(\lam_{231}) &=& \frac{\lam_{231}}{16\pi^2} \biggl [
Y_\tau^2+4 \lam_{231}^2+\lam_{123}^2+\lam_{132}^2 -
\biggl (\frac{9}{5} g_1^2+3 g_2^2 \biggr ) \biggr ] \\
\beta(\lam_{132}) &=& \frac{\lam_{132}}{16\pi^2} \biggl [
Y_\tau^2+4 \lam_{132}^2+\lam_{123}^2+\lam_{231}^2 -
\biggl (\frac{9}{5} g_1^2+3 g_2^2 \biggr ) \biggr ] 
\end{eqnarray}

\subsection{Other soft terms}
The soft terms for squark masses and trilinear couplings
can be derived from the general formulae
in the Appendix. 
To one-loop accuracy, the rest are equivalent to the $R_p$
conserving MSSM soft terms~\cite{wells}, except for the trilinear slepton
couplings and $m_{H_1}$:
\begin{equation}
m_{H_1}^2 = \frac{M_{aux}^2}{16\pi^2} \biggl [ 3 Y_b \beta(Y_b) +  Y_\tau
\beta(Y_\tau)
-\biggl ( \frac{3}{10} g_1 \beta(g_1) + \frac{3}{2}g_2 \beta(g_2) \biggr )
\biggr ].
\end{equation}
From eq.~(\ref{ytaui}), we see that $m_{H_1}$ depends upon the combination
$\kappa \equiv 2 \lam_{123}^2 + \lam_{132}^2+\lam_{231}^2$. 
Because $\mu$ is fixed partly by
$m_{H_1}$ in the electroweak symmetry breaking condition~\cite{BandB}, it is
altered from 
the $R_p$-conserving scenario by $\kappa \neq 0$. 
We note in particular the anomaly-mediated
contribution to the $B$-term realised in a specific model with a bulk
contribution~\cite{rs}:
\begin{eqnarray}
B=-\frac{M_{aux}}{16\pi^2} \biggl [ 3 Y_t^2 +3 Y_b^2+Y_\tau^2 -
\biggl (\frac{3}{5} g_1^2 +3 g_2^2 \biggr ) \biggr ]. \label{Bterm}
\end{eqnarray}
We shall utilise this model in order to cut the parameter space down.
The prediction of $B$ (for a given value of $\maux$) results in a prediction
of $\tan \beta$ 
from the potential minimisation conditions~\cite{BandB}. 
We note that in
the AMSB MSSM, a term $\mu H_1 H_2$ in the superpotential spoils the conformal
invariance. However, $\mu$ can be viewed as a result of supersymmetry
breaking~\cite{rs}, providing a natural explanation the size of $\mu$
necessary to obtain $M_Z=91.18$ GeV. 
In our convention, $B$ and $\mu$ have
opposite signs in successful minima, so the $B$ term predicted also constrains
the sign of $\mu$ to be positive.

\section{Spectra and constraints}

\label{sec:cons}
Some products of the \lfv-couplings are constrained to be tiny and practically
useless in solving the AMSB slepton mass problem~\cite{second}, 
\begin{eqnarray}
\lambda_{mni} &\ntw& \lambda_{mnj} \;\;\;  \;\;\; i\neq j ,
\label{s3}\\
\lambda_{imn} &\ntw& \lambda_{jmn} \;\;\;  \;\;\; i\neq j ,
\label{s4}
\end{eqnarray}
where $\ntw$ stands for `not non-zero with'.
The combination of couplings $\lambda_{123}, \lambda_{132},
 \lambda_{231} \neq 0$ respects this constraint.
In addition, the most recent bounds upon the individual couplings
are~\cite{uslot}: 
\begin{eqnarray}
\lambda_{123}
&\lsim  & 0.49\times \frac{m_{\tilde{\tau}_R}}{1~{\mathrm TeV}}\label{bound} \\
\lambda_{132}&\lsim & 0.62\times \frac{m_{\tilde{\mu}_R}}{1~{\mathrm TeV}} \\
\lambda_{231}&\lsim & 0.70\times \frac{m_{\tilde{e}_R}}{1~{\mathrm TeV}}, \label{bound2}
\end{eqnarray}
from charged lepton universality~\cite{BGH}. 
As we show below, these provide the most severe constraints upon the model.
The couplings also pass the $\mu
\rightarrow e \gamma$ conversion limits~\cite{muegam}.

\subsection{Sparticle Spectra}

\FIGURE[t]{\psfig{figure=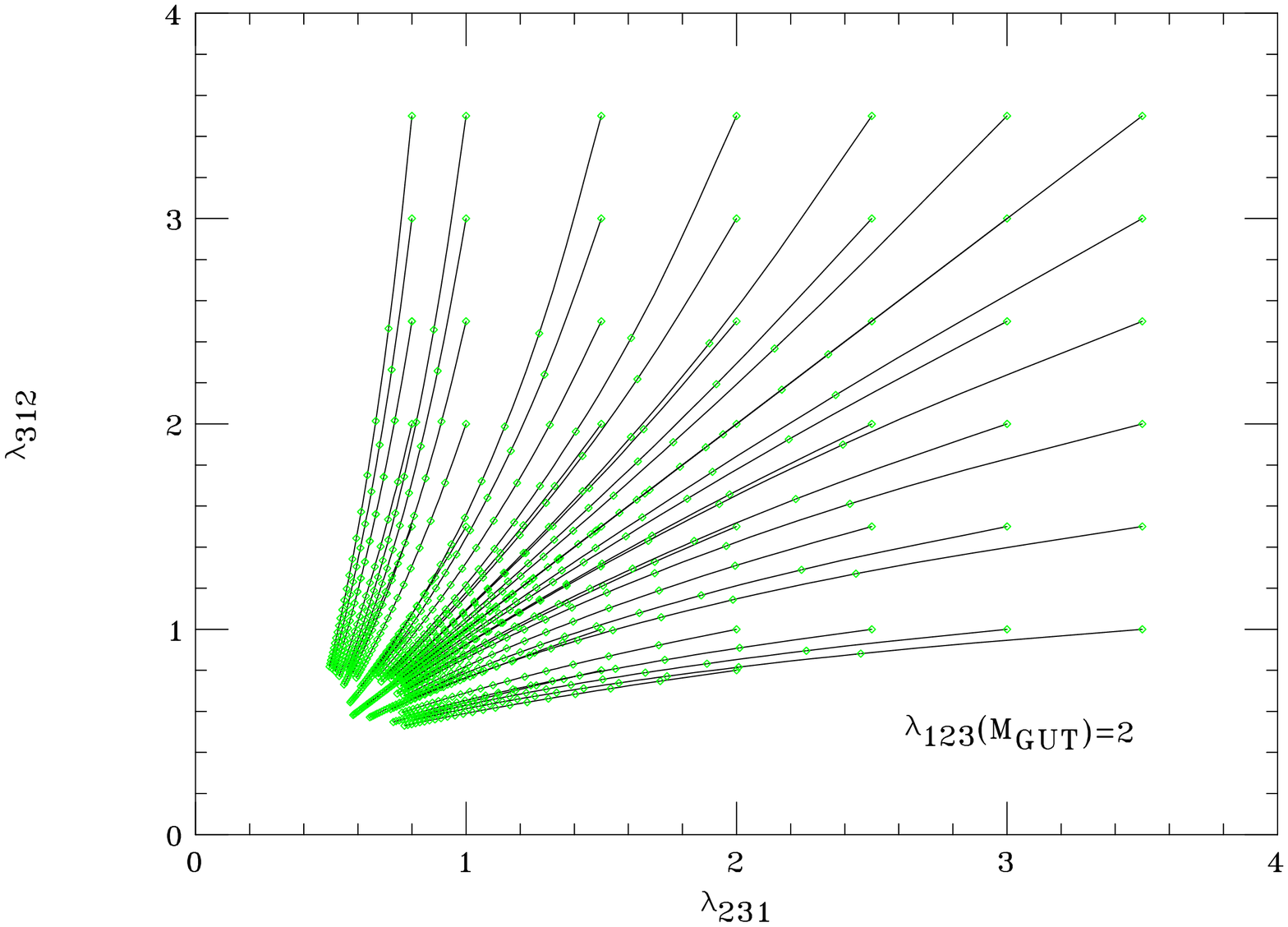,height=4in,angle=90}
\label{fig:quas}
\caption{Quasi-fixed structure of the \rpv-couplings. Renormalisation group
trajectories of the couplings $\lam_{132},\lam_{231}$ are shown for
various boundary conditions at $M_{GUT}$. We display the running from
$M_{GUT}$ to the
electroweak scale, where all couplings approach the $\lam_{132} \sim 0.75$,
$\lam_{231}
\sim 0.75$ region. The dots denote a decrease in the renormalisation scale by
a factor of 100.
}}
We now perform a one-loop accuracy numerical analysis to determine the
sparticle and Higgs spectrum. The full one-loop Higgs potential is minimised at a
scale $Q \equiv $2 TeV, where the radiative corrections to the potential are
small, to
determine $\mu$. This choice of scale can change the $\tan \beta $ prediction
a little.
For experimental inputs on 
the gauge couplings we use~\cite{PDB} $\alpha_s(M_Z)=0.119$, $\sin^2
\theta_w=0.2312$, $\alpha(M_Z)=1/127.9$ in the $\overline{MS}$ scheme.
3 loop QCD$\otimes$2 loop QED is used as an effective field theory below
$M_Z$. Central values~\cite{PDB} of the top and bottom pole masses
$M_t=174.3, M_b=4.9$ GeV are taken. For simplicity we set
$\lam \equiv\lam_{123}=\lam_{231}=\lam_{312}$, a renormalisation group
invariant
choice in the limit that $Y_\tau=0$. In fact, all soft masses and couplings
except 
those of the sleptons and one of the Higgs are independent of the choice of
these three couplings. 
Here, we choose $\lam(M_Z)=0.73$, which is sufficient to render all slepton
mass
squared values positive ($\lam > 0.66$). If $\lam(M_Z)$ is set too large, then
$M_{aux}$ must be set very large in order to produce sleptons heavy enough to
evade eqs.~(\ref{bound}-\ref{bound2}). Remarkably, $\lam(M_Z) = 0.73$ is near
a common
quasi-fixed point value for all three couplings if we assume that they are set
large at a scale $M_{GUT} \sim 2 \times 10^{16}$ GeV, where the gauge couplings
unify. The quasi-fixed behaviour is
exhibited by displaying insensitivity to the ultra-violet boundary
condition~\cite{quasi}. 
This behaviour is exhibited in Fig.~\ref{fig:quas}
\FIGURE[t]{\psfig{figure=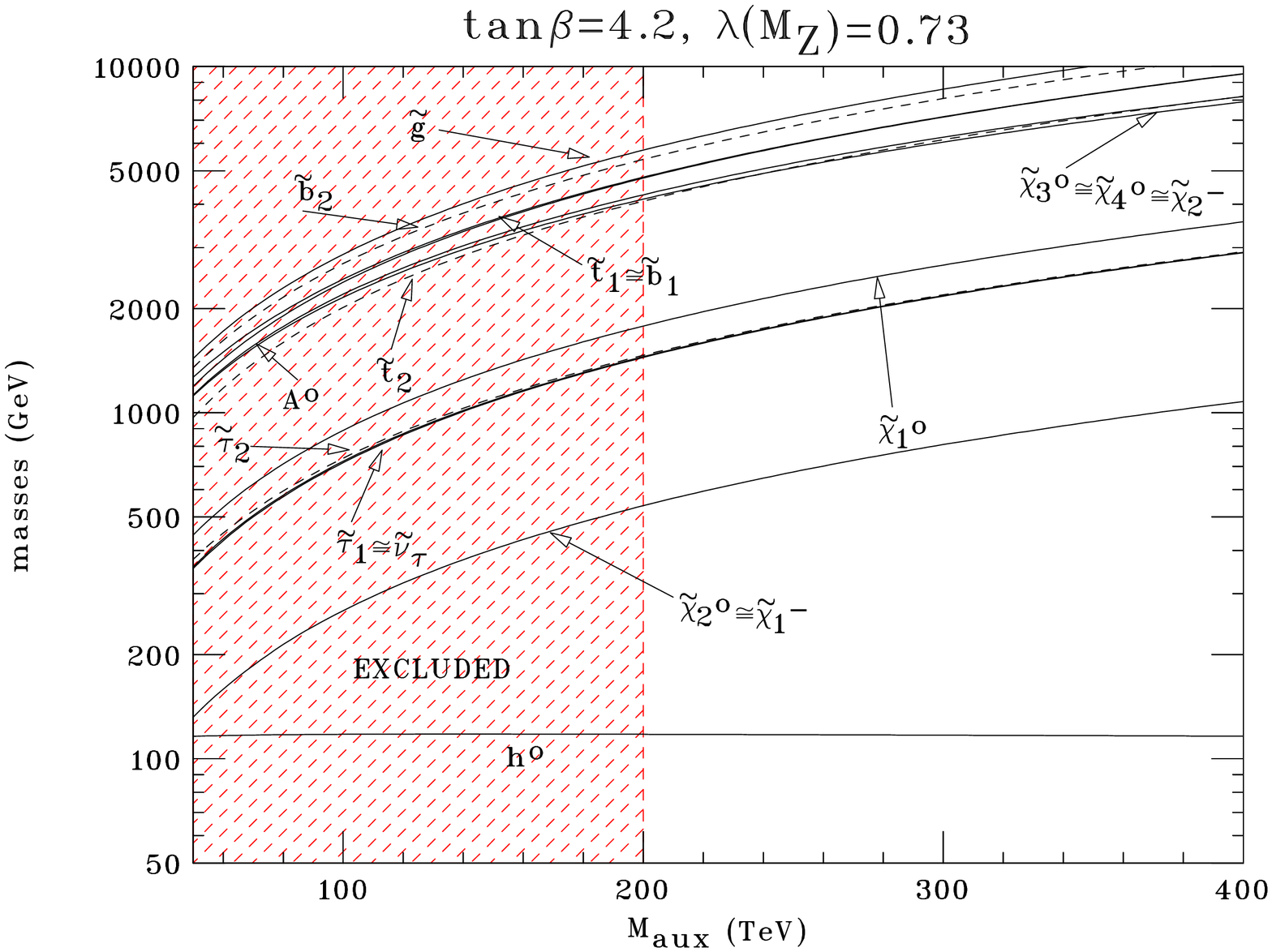,height=4.5in,angle=90}
\label{fig:spectrum}
\caption{The supersymmetric particle spectrum in our AMSB scenario
for different values of $M_{aux}$.  The dashed region is excluded from
charged lepton universality constraints.}}
As can be seen from the figure, $\lam_{132}$ and $\lam_{231}$ both approach
the 0.7-0.8 region, roughly independent of the values assumed for them at
$M_{GUT}$. We have checked that $\lam_{123}$ approaches the 0.7-0.8
region also.

Fig.~\ref{fig:spectrum} shows the supersymmetric particle spectrum in the AMSB
scenario
for different values of $M_{aux}$. The value of $\tan\beta\approx 4.2$
is predicted where the minimisation conditions of the Higgs potential
satisfy eq.~(\ref{Bterm}). In fact, this value has a small (neglected)
dependence
upon $M_{aux}$ and the scale at which the potential is minimised, and thus has
an uncertainty of about $\pm 1.0$.
The dashed region is excluded from the
experimental bounds derived from charged lepton universality, the most
stringent being eq.~(\ref{bound}). An improvement of these bounds
by a factor of two would test up to $M_{aux}=400$ TeV.
The LSP neutralino 
is quasi-degenerate  with the lightest chargino, as usual in
AMSB~\cite{wells}. 
The Higgs mass determinations are performed using state-of-the-art two loop
corrections~\cite{svenandgeorg}. 
The lightest CP-even Higgs mass $m_{h^0}$ is insensitive to $\maux$ and
$\lambda(M_Z)$, but has the usual large dependence upon $M_t$. For $M_t=174.3$
GeV however, $M_{h^0}=117.5 \pm 0.5$ GeV, with an estimated 2 GeV uncertainty
coming from higher order corrections.
As can be seen from the figure, the excluded region forces all sparticles to
be rather heavy - the lightest chargino and neutralino can be as light as 500
GeV, but all other sparticles and heavy Higgs must be heavier than 1.1 TeV.

\section{Collider phenomenology}
\label{sec:coll}

\TABULAR[t]{ccccc}
{\label{tab:one}
$A_t(M_Z)=$6.53  &
$A_b(M_Z)=$11.97  &
$A_\tau(M_Z)=$-0.71  &
$B(M_Z)=$-1.42  &
$\mu(M_Z)=$4.53  \\
$m_{\tilde{g}}=$6.31  &
$m_{\tilde{\chi}_1^0}=$1.96  &
$m_{\tilde{\chi}_2^0}=$0.59 & 
$m_{\tilde{\chi}_3^0}=$4.53 & 
$m_{\tilde{\chi}_4^0}=$4.53  \\
$m_h=$0.1176 &
$m_A \approx m_H=$4.67  &
$M_{H^\pm}=$4.67  &
$m_{\tilde{\chi}_1^\pm}=$0.59  &
$m_{\tilde{\chi}_2^\pm}=$4.53  \\
$m_{\tilde{\tau}_{1,2}}=$1.60  &
$m_{\tilde{\nu}_\tau}=$1.60  &
$m_{\tilde{e}_1,\tilde{\mu}_1}=$1.60  &
$m_{\tilde{e}_2,\tilde{\mu}_2}=$1.61  &
$m_{\tilde{\nu}_e,\tilde{\nu}_\mu}=$1.60  \\
$m_{\tilde{t}_1}=$5.27  &
$m_{\tilde{t}_2}=$4.50  &
$m_{\tilde{b}_1}=$5.24  &
$m_{\tilde{b}_2}=$5.93  & \\
$m_{\tilde{u}_1,\tilde{c}_1}=$5.86  &
$m_{\tilde{u}_2,\tilde{c}_2}=$5.92  &
$m_{\tilde{d}_1,\tilde{s}_1}=$5.86  & 
$m_{\tilde{d}_2,\tilde{s}_2}=$5.94  & \\
}
{Spectrum and couplings for $\maux=220$ TeV and $\lam(M_Z)=0.73$.
Masses are measured in TeV.}
\TABULAR[t]{c|cc}
{\label{tab:cross} 
& $\sigma$(Pb) & No.\ events\\ \hline
\multicolumn{3}{c}{LHC} \\ \hline
$\chi_i^{\pm,0}$ &  0.025 & 720 \\
$\tilde{q},\tilde{g}$ & $10^{-11}$ & 0 \\
$\tilde{l},\tilde{\nu}$ & $6\times 10^{-6}$ & 0 \\ \hline
\multicolumn{3}{c}{Tevatron} \\ \hline
All SUSY & $2.3\times 10^{-6}$ & 0 \\
}{Cross sections $\sigma$ of sparticle production at the Tevatron and LHC.}
The phenomenology and search prospects of the AMSB $R_p$-conserving scenarios
have been considered in
refs.~\cite{rs},\cite{degenDKs},\cite{feng},\cite{wells},\cite{AMSBph}. The
present scenario differs in two main respects. Firstly, the \rpv-coupling
exclusion limits shown in fig.~\ref{fig:spectrum} force superpartners to be
heavier than was previously considered with $R_p$-conserving AMSB\@. Secondly,
the decays of $\chi_1^\pm$, $\chi_1^0$ are qualitatively different. In the
$R_p$ conserving case, the quasi-degeneracy of $\chi_1^\pm$ and $\chi_1^0$
means that the $\chi_1^\pm$ is quasi-stable~\cite{degenDKs}, and of course the
$\chi_1^0$ is undetected, except as missing energy. A classic signature for the
lightest chargino is then the presence of heavily ionising tracks, with
possible slow decays into pions/leptons. In the present scenario however, the
lightest chargino and neutralinos decay almost immediately into 3 leptons.

To illustrate the decays and cross sections of the model, we pick a particular
value of $\maux=220$ TeV. The detailed spectrum and parameters are displayed
in Table~\ref{tab:one}. {\small HERWIG6.1}~\cite{HERWIG} was utilised to
estimate sparticle discovery prospects of this
spectrum at the Tevatron and LHC\@.
We display the cross-sections of the hard sub-process of sparticle
production in Table~\ref{tab:cross}. Also shown is the number of expected
events for luminosities of ${\mathcal L}=10,30$ fb$^{-1}$ at the Tevatron and
LHC respectively\footnote{Equivalent to approximately 3 years running at low
luminosity at the LHC}. The table shows that charginos and neutralinos are
produced at the LHC at a detectable rate, but the Tevatron should see no
superparticles. However, we note that the lightest CP-even Higgs of mass
$m_{h^0}\approx 118$ GeV should be discovered at the Tevatron~\cite{tevhiggs}.

We ran the weak-scale spectrum through a version of {\small
ISASUSY}~\cite{ISAJET} modified to take \rpv-interactions into
account~\cite{prich}. This then calculated the relevant decays.
The lightest neutralino decays through twelve channels with equal branching
ratios of 1/12 and partial widths of 2.25$\times 10^{-5}$ GeV:
\begin{eqnarray}
\chi_1^0 &\rightarrow& \left. e^{+}\bar \nu_\mu \tau^-, \ e^- \nu_\mu
\tau^+, \ e^+ \bar \nu_\tau \mu^-, \ e^- \nu_\tau \mu^+, \
\mu^{+}\bar \nu_e \tau^-, \ \mu^- \nu_e \tau^+, \right. \nonumber \\
&& \left. 
\mu^+ \bar \nu_\tau e^-, \ \mu^- \nu_\tau e^+, \
\tau^{+}\bar \nu_e \mu^-, \ \tau^- \nu_e
\mu^+, \ \tau^+ \bar \nu_\mu e^-, \ \tau^- \nu_\mu e^+
 \right.,
\end{eqnarray}
whereas $\chi_1^+$ has six decay channels
\begin{equation}
\chi_1^+ \rightarrow \nu_\mu \nu_e \tau^+, \ 
\nu_\tau \nu_e \mu^+, \
\nu_\tau \nu_\mu e^+, \
\mu^+ e^+ \tau^-, \
\tau^+ e^+ \mu^-, \
\tau^+ \mu^+ e^-
\end{equation}
again with equal branching ratios of 1/6 and partial widths of 4.5$\times
10^{-5}$ GeV. These decays should be easy to find with low backgrounds at the
LHC\@. Double chargino production can be found by decays into six charged
leptons, or chargino/neutralino production via a five charged lepton channel,
with distinctive flavour structure. Lepton flavour violation is usually
explicit in the final state. We have obtained approximately equal branching
ratios here mainly because we have assumed
$\lam_{123}=\lam_{231}=\lam_{132}$. In the case they are non-degenerate, this
will change and the relative branching ratios into different final states will
help measure the \rpv-couplings.

\section{Conclusions}
\label{sec:conc}

We have proposed a new solution to the problem of slepton negative mass
squared values in the AMSB MSSM\@. It involves including 3 \lfv-operators in
the
superpotential which were previously assumed to be absent. This leads to
positive mass squared values for all of the sleptons and renormalisation-group
invariant relations between supersymmetry breaking terms and the measured
supersymmetric couplings. This has the advantage of rendering the model
insensitive to unknown ultra-violet effects. The $\mu$ problem has a natural
solution~\cite{rs}, indeed the prediction of the $B-$term in a particular
model results in a
prediction for $\tan \beta$. All of the sparticle spectrum except for the
slepton masses are then given in terms of two free parameters:
$\maux$ and $\kappa$.

We have therefore assumed the MSSM spectrum near the weak scale, and that the
dominant source of supersymmetry breaking terms in the observable sector is
from anomaly mediation.
Experimental limits on the \lfv-operators provide stringent constraints upon
the model, meaning that sparticle masses must be rather high. Whereas the
lightest charginos and neutralinos can be as light as 500 GeV, the other
sparticles must be heavier than 1.1 TeV. The Tevatron therefore sees no new
particles except the
lightest Higgs of mass 118 GeV, and the LHC can detect the  lightest charginos
and neutralinos via distinctive leptonic decays. Charged lepton universality
violation 
is predicted to be close to its experimental bound, within a factor of two.

Neutrino masses and mixings are beyond the scope of this paper, but it is well
known that the \lfv-operators we have introduced can generate them at the loop
level~\cite{RPVneut}. We intend to pursue them in future work~\cite{fut}, and
the small number of free parameters should allow a strict correlation with
lepton flavour violating predictions. We believe the present model of
supersymmetry breaking in the observable sector to warrant several future
works. For example, a more accurate calculation of the spectrum and a
determination of the LHC reach in parameter space would be useful. It would be
desirable to find symmetries to ban or suppress the other \rpv-couplings. 
Aside from these, the usual calculations in the MSSM (quark FCNCs, charge and
colour breaking constraints etc) could be performed. It will be interesting to
investigate the present idea in a more general framework, for example when
$\tan 
\beta$ is large (the prediction of the lightest MSSM Higgs mass is likely to
change from the one presented here) and there are only two LLE couplings, or
where splittings
between the \rpv couplings occur. Relaxing the assumption that $\lam'_{ijk} =
0$ might lead to the possible observation of a single slepton at the LHC via
slepton-strahlung~\cite{Borzumati:1999th}. 

To summarise, our scenario is a predictive scheme of supersymmetry breaking,
containing natural solutions to the $\mu$ problem and supersymmetric flavour
problem. The spectrum depends upon only two parameters apart from the slepton
masses. In the case that the
\rpv-couplings are at their quasi-fixed point values, the slepton masses
approximately only depend upon these same two parameters.
If one assumes a high-energy cut-off scale, such as the GUT scale for example,
we note that the weak-scale values of the couplings are approximately
predicted by the quasi-fixed point and there is only one free parameter on
which the whole sparticle spectrum depends.
$\tan \beta$ is predicted in a specific model and the soft masses are given by
renormalisation group invariant relations with the measured SUSY couplings.
The phenomenology is rather distinctive and should be
easily disentangled from other possibilities at the LHC, after being tested at
the Tevatron by
the Higgs mass prediction. The present model is the only
example of a 
model with both insensitivity of the soft terms to unknown ultra-violet
physics and the MSSM spectrum near the weak scale, and as such is important to
investigate further.

\begin{appendix}

\section{One-loop anomalous dimensions and beta functions in the \rpv-MSSM}

$\lamu{i},\lamd{i},\lame{i}$ were written in a matrix notation in~\cite{uslot}
as
\begin{eqnarray}
(\lame{k})_{ij}=\lambda_{ijk} ,\qquad
(\lamd{k})_{ij}=\lambda^\prime_{ijk}, \qquad
(\lamu{k})_{ij}=\lambda_{ijk}^{\prime \prime}
\end{eqnarray}
and we adopt this notation for presenting results with general family
structures. The one-loop anomalous dimensions of the \rpv-MSSM are~\cite{uslot}
\begin{eqnarray}
16\pi^2~\Gamma^{(1)L_j}_{L_i} &=&\left(\ye \ye^\dagg \right)_{ji}
+(\lame{q}\lame{q}^\dagg)_{ji} +3 (\lamd{q}\lamd{q}^\dagg)_{ji}
-\delta_i^j(\frac{3}{10}g_1^2+\frac{3}{2}g_2^2), \label{gamml}\\
16 \pi^2~\Gamma^{(1)E_j}_{E_i} &=& 2 \left(\ye^\dagg \ye \right)_{ji}
+ \mbox{Tr}(\lame{j}\lame{i}^\dagg) -\delta_i^j(\frac{6}{5}g_1^2),
\label{gamme}\\
16 \pi^2~\Gamma^{(1)Q_j}_{Q_i} &=& \left(\yd \yd^\dagg \right)_{ji}
+ \left(\yu \yu^\dagg \right)_{ji}
+ (\lamd{q}^\dagg\lamd{q})_{ij}  \nonumber \\
&& -\delta_i^j(\frac{1}{30}g_1^2+\frac{3}{2}g_2^2+\frac{8}{3}g_3^2),\\
16 \pi^2~\Gamma^{(1)D_j}_{D_i} &=& 2 \left(\yd^\dagg \yd \right)_{ij}
+2 \mbox{Tr}(\lamd{i}^\dagg\lamd{j}) +2 (\lamu{q}\lamu{q}^\dagg)_{ji}
\nonumber \\ &&
-\delta_i^j(\frac{2}{15}g_1^2+\frac{8}{3}g_3^2)),\\
16 \pi^2~\Gamma^{(1)U_j}_{U_i} &=& 2\left(\yu^\dagg \yu \right)_{ij}
+ \mbox{Tr}(\lamu{j}\lamu{i}^\dagg)
-\delta_i^j(\frac{8}{15}g_1^2+\frac{8}{3}g_3^2)),\\
16 \pi^2~\Gamma^{(1)H_1}_{H_1} &=& \mbox{Tr}\left(3\yd\yd^\dagg+\ye\ye^\dagg
 \right)
-(\frac{3}{10}g_1^2+\frac{3}{2}g_2^2),\\
16 \pi^2~\Gamma^{(1)H_2}_{H_2} &=& 3 \mbox{Tr}\left( \yu\yu^\dagg\right)
-(\frac{3}{10}g_1^2+\frac{3}{2}g_2^2),\\
16 \pi^2~\Gamma^{(1)H_1}_{L_i} &=& {16 \pi^2~\Gamma^{(1)L_i}_{H_1}}^* 
=-3 (\lamd{q}^*\yd)_{iq}
- (\lame{q}^*\ye)_{iq}. \label{LHmix}
\end{eqnarray}

The beta functions of the couplings appearing in the
superpotential in eq.~(\ref{superpot}) are:
\begin{eqnarray}
 \beta({\ye})_{ij} &=& (\ye)_{ik}\ggam{E_j}{E_k}
+(\ye)_{ij}\ggam{H_1}{H_1}
-(\lame{j})_{ki}\ggam{H_1}{L_k}
+(\ye)_{kj}\ggam{L_i}{L_k},\label{eq:ye}\\[0.2cm]
 \beta({\yd})_{ij} &=& (\yd)_{ik}\ggam{D_j}{D_k}
  +(\yd)_{ij}\ggam{H_1}{H_1}
  -(\lamd{j})_{ki}\ggam{H_1}{L_k}
  +(\yd)_{kj}\ggam{Q_i}{Q_k},\label{eq:yd}\\[0.2cm]
 \beta({\yu})_{ij} &=& (\yu)_{ik}\ggam{U_j}{U_k}\label{yu}
+(\yu)_{ij}\ggam{H_2}{H_2}
+(\yu)_{kj}\ggam{Q_i}{Q_k}, 
\\[0.2cm]
\beta ({\lame{k}})_{ij} &=& (\lame{l})_{ij}\ggam{E_k}{E_l}
+(\lame{k})_{il}\ggam{L_j}{L_l}
+(\ye)_{ik}\ggam{L_j}{H_1} \nonumber \\ &&
-(\lame{k})_{jl}\ggam{L_i}{L_l}-(\ye)_{jk}\ggam{L_i}{H_1}, 
\label{eq:lame}\\[0.2cm]
\beta({\lamd{k}})_{ij} &=& (\lamd{l})_{ij}\ggam{D_k}{D_l}\lab{lamd}
+(\lamd{k})_{il}\ggam{Q_j}{Q_l}
+(\lamd{k})_{lj}\ggam{L_i}{L_l}
-(\yd)_{jk}\ggam{L_i}{H_1}, \\[0.2cm]
\beta (\lamu{i})_{jk} &=& (\lamu{i})_{jl}\ggam{D_k}{D_l} \lab{lamu}
+(\lamu{i})_{lk}\ggam{D_j}{D_l}
+(\lamu{l})_{jk}\ggam{U_i}{U_l}.
\end{eqnarray}

\end{appendix}

\acknowledgments
BCA would like to thank D.R.T. Jones, P. Richardson and M.A. Parker for useful
discussions and the IFIC, University of Valencia for hospitality offered while
some of this work was carried out. We thank H. Dreiner for useful
discussions and comments on the draft.
AD is supported from Marie Curie Research Training Grants
ERB-FMBI-CT98-3438 and thanks DAMTP, University of Cambridge for hospitality.
This work has been partially supported by PPARC.

\end{document}